\documentstyle[12pt]{article}
\newcommand{\bb}{\begin{equation}}   \newcommand{\ee}{\end{equation}}
\newcommand{\beqa}{\begin{eqnarray}} \newcommand{\eeqa}{\end{eqnarray}}
  
  \newcommand{\as}{\alpha_{s}}

\title{\begin{flushright}
{\normalsize TPI--MINN--95--17/T; NUC--MINN--95--15/T; HEP--UMN--TH--1348\\
hep-lat/9505025; IPS Report 95-15; MSI Report UMSI 95/238\\
Submitted to Nucl. Phys. B ; Revised version, January 1996\\}
        \end{flushright}
{\bf Determination of the renormalized\\
 heavy-quark mass in Lattice QCD \\}}
\author{{\bf A. Bochkarev}$^{a, \dagger}$ and 
{\bf Ph. de Forcrand}$^{b}$ \\
{\it $^a$ TPI, U. of Minnesota, Minneapolis, MN 55455, USA}\\
{\it $^b$ IPS, ETH-Zentrum, Z\"urich CH-8092, Switzerland}}

\date{}

\begin{document}

\maketitle
\begin{center}
Abstract\\
\end{center}
We study on the lattice the correlator of heavy-quark currents in the
vicinity of vanishing momentum. 
The renormalized charmed quark mass, the renormalized strong coupling 
constant and gluon condensate can be defined in terms of the derivatives
of that correlator at zero momentum. We analyze quenched Monte-Carlo 
data on a small lattice $8^3*16$ for $\beta=6$. We generalize 
dispersion relations to the lattice theory in a simple way and use them 
successfully to fit the correlator at both small and large distances. 
We fit the short-distance part of the correlator with the relevant 
expressions of perturbative QCD on the lattice and obtain the value
of the renormalized quark mass $m_c^{\bar{MS}}(m_c)\,=\,1.20(4)\,GeV$.\\

$^\dagger$ On leave from: INR, Russian Academy of Sciences,
Moscow 117312, Russia

\vfill \eject
\section*{Introduction}
At present the accurate calculations of the strong coupling constant $\alpha_s$ 
\cite{bethke} and the masses of heavy quarks \cite{nar}, \cite{ynd} are not 
only the subject of intensive studies in the continuum QCD. Monte-Carlo
Lattice simulations seem to provide us with competitive ways to evaluate
$\alpha_s$ \cite{fermilab},\cite{davies} and heavy-quark masses \cite{ukqcd} with 
considerable accuracy.
A systematic study of the correlator of heavy-quark currents $\Pi(q^2)$
near zero momentum $q^2 = 0$ on the lattice was suggested recently
\cite{b} as a way to calculate the renormalized parameters of perturbative
QCD, the strong coupling constant $\alpha_{s}$ and heavy quark mass $m_{c}$, 
as well as the main parameter of non-perturbative QCD, the vacuum gluon 
condensate 
$G^{(2)}\,\equiv\,<0|(\alpha_{s}/\pi)\,G_{\mu\nu}^{a}G_{\mu\nu}^{a}|0>$ 
\cite{svz1}.

In this paper we formulate the method in more detailes and analyze Monte-Carlo
data on relatively small lattice $8^3*16$, as a {\em pilot} study. We 
concentrate on the renormalized mass of the charm quark $m_c$. The renormalized parameters of the QCD Lagrangian, $\alpha_{s}(\mu^2)$ and
$m_{c}(\mu^2)$ serve as the boundary conditions to the renormalization group
equations. Here $\mu^2$ is some physical normalization point, which
stays finite as the cut-off is removed, say the lattice spacing 
$a \rightarrow 0$. This choice of a physical normalization point allows one 
to easily make use of phenomenological data in the determination of 
$\Lambda_{QCD}$ and current quark masses on the Lattice. It also
helps to isolate such non-perturbative quantities as the gluon condensate,
whose naive definition would imply a strong power-like dependence on the
normalization point \cite{svz2}. This is the main difference between the
proposal \cite{b} and the works \cite{dig} where the computation of the
gluon condensate on the lattice was pioneered. 

Because of the {\em asymptotic freedom} the boundary conditions to the 
renormalization group equations in
QCD should be imposed at large Euclidean momenta corresponding to small
space-like distances. The renormalized parameters of QCD are essentially the
short-distance quantities. We would like to emphasize the difference between 
the pole-mass of quarks and Euclidean quark mass. The pole mass is a parameter
of the {on-shell}-subtraction scheme: the ultraviolet singularities are
subtracted on the quark mass-shell. This scheme does not appeal because 
of the complicated infrared properties of QCD: quarks do not exist on the
mass-shell. The pole mass of quarks is well defined only within the first orders
of perturbation theory, while the Euclidean mass is defined in the domain 
where perturbative QCD applies. In the popular these days
language one would say that the pole-mass suffers from the infrared-renormalon
ambiguities while the Euclidean mass does not.  

One way to calculate the Euclidean mass on the lattice was considered in 
\cite{gym}. The prescription is: $a)$ within the lattice theory calculate the 
pole mass $m^{\ast}$ of the quark propagator as a function of the 
parameters of the lattice theory $m^{\ast}[\kappa, r, \beta,...]$  
perturbatively in powers of the inverse bare 
coupling constant $\beta$. Here : $\beta=6/g^2$ and $\kappa, r$ are 
parameters of the Wilson propagator. This perturbative expansion is good at 
large $\beta$, where the asymptotic scaling holds: 
the function $g^2(a)$ solves the perturbative renormalization group equation. 
Volume has been always considered infinite in these analytic calculations. 
The essential ingredient of this calculation is the bare quark mass $m(a)$ ,
which is obtained by fitting the 
Monte-Carlo data for the pion mass $m_{\pi}$ with the ansatz: 
$m(a)\,=\,(1/\kappa\,-\,1/\kappa_{cr})/2$.
The motivation for this fit is that $m^{2}_{\pi}\sim m_{quark}$ is 
expected in the chiral limt. The fit is done for $\kappa$s corresponding to
light fermions in the finite volume and for relatively small $\beta$s,
where $\beta$ does not scale.
$b)$ Use the relation of the continuum theory between the quark pole mass and
the Euclidean mass $m^{\ast}[m(\mu),\alpha_s (\mu)]$.

Our approach is somewhat alternative to $\cite{gym}$. We use the correlator 
of hadronic currents of heavy quarks to determine the Euclidean quark mass. 
We never refer to the pole mass. The correlator of the heavy-quark currents
is reliably calculated in perturbative QCD at vanishing momentum, because
that point is away from the nearest threshold due to heavy quark-antiquark
pair production. The heavy-quark threshold $4m_{c}^{2}$ is the relevant
virtuality here. The perturbative loop-expansion is built in powers of small
parameter $\as(4m_{c}^{2})$. {\em One} loop of free quarks with the 
renormalized Euclidean mass gives a very good approximation to the 
correlator of the charm-quark currents near vanishing momentum \cite{svz1}. 
This is in contrast to the correlator of light quarks, which is
not analytic at vanishing momentum and cannot be computed at that point 
within perturbative QCD. 

The correlator of the gauge-invariant local hadronic currents $\Pi(x)$ is a 
very conventional thing to study in Lattice QCD. The long-distance behavior
of the correlator is determined by the low-lying resonance. To obtain
the resonance mass one fits the zero-momentum correlator $\Pi(t)$ by 
$cosh[(t-L/2)m_{res}]$, which describes propagation of a single particle of 
mass $m_{res}$ in a finite volume $L$ in the continuum theory ($a=0$). 
We do the same thing: fit the long-distance part of the correlator of
heavy-quark currents with the single-resonance approximation to that
correlator. This allows us to fix the scale of our Monte-Carlo data. In
addition to that we study the short-distance part of the correlator as well. 
We fit it with the corresponding expressions of perturbative QCD
in order to extract the fundamental parameters of perturbative QCD: the
renormalized quark mass and the strong coupling constant. 

In order to separate the short-distance part of the correlator from its
long-distance part we study derivatives of the correlator $\Pi(q^2)$ with
respect to momentum $q^2$ at the origin $q^2=0$, called moments 
\cite{svz1}. The lower derivatives probe the vicinity of the origin, which is
$4m_{c}^{2}$ $GeV^2$ away from the charm quark-antiquark threshold, hence 
it is in the domain of aymptotic freedom. 
The lower moments are reliably computed in perturbative QCD. 
We use Monte-Carlo data for the lower moments to extract the parameters
of perturbative QCD. Higher moments probe the values of $q^2$ away from the 
origin. They represent the long-distance part of the correlator and are 
entirely saturated by the low-lying resonance. We use those moments 
in order to fix the scale of the lattice theory.

The next section 1.1 is a short introduction into the so-called dispersion theory
of charmonium \cite{svz2}. We describe analytic properties of the two-point 
correlator of charm-quark currents in the continuum theory and explain why it 
is reliably calculated near vanishing momentum in perturbative QCD. The other
sections are our attempt to extend the dispersion theory of charmonium to the
Lattice QCD. 

In section 1.2 we discuss an extension of the dispersion relations to the
lattice theory.
Since we deal with the lattice theory of heavy quarks significant cutoff effects
 are expected. At $\beta = 6$ the lattice spacing $a\;\sim\;(2\,GeV)^{-1}$.
For the charm quark mass $m_c\approx 1.3 GeV$ one has the parameter of the 
deviation of the fermionic action from the continuum $a\cdot m_c\,\approx\,0.6$
 - moderately, but not comfortably small. Actually one would like to probe
virtualities $\mu^2\,\sim \,4 m_{c}^2$ on the lattice, then 
$a\cdot \mu\,\sim\,O(1)$. From the phenomenological point of view we are trying
to generate a resonance of mass $\sim\;m_{J/\psi}\simeq 3.1 GeV$. Then the resonance
mass in the units of the lattice spacing $a\cdot m_{res} \,\approx\,1.5$. Such
lattice theory is obviously coarse enough to exhibite certain cutoff effects in
the propagation of a single resonance. We explore the cutoff effects via fitting
Monte-Carlo data with the following generalization of the dispersion relation 
for the two-point correlator $\Pi (q^2)$ to the lattice theory:
\begin{equation}
\int ds\; \frac{\rho(s)}{s\;+\;q^2}\;\;\;\Longrightarrow \;\;\;\int ds\; 
\frac{\rho(s, a, L)}{s\;+\;\frac{4}{a^2} \,\sum_{\mu}^{4} \sin^{2}
(q_{\mu} a/2)}  \label{disp1}
\end{equation}
where $L$ represents the finite volume of the box, and the function $\rho(s, a, L)$
approaches the corresponding spectral density of the continuum theory in the
limit $a\rightarrow0,\;L\rightarrow\infty$. We find significant evidence
that although the l.h.s. of (\ref{disp1}) exhibites strong deviations due
to  $a\neq0\,\&\,L\ll\infty$, the dependence of the spectral density on $a$
and $L$ is rather weak on the lattice under consideration. In other words, most
of the cutoff effects in the heavy-quark systems are successfully taken into
account by the naive modification of the denominator in (\ref{disp1}).

In section 2 we compare the moments of the correlator in a single-resonance 
approximation on the finite lattice and in the continuum theory. We compare the
fit based on the dispersion relation (\ref{disp1}) with the traditional fit to
$cosh[(x-L/2) m_{res}]$ for different values of the parameter $a\cdot m_{res}$.
We analyze Monte-Carlo data obtained on a blocked lattice $8^3\times 16$ with 
$a^{-1} \simeq 500 MeV$, which is supposed to show correctly the long-distance part
of the correlator, where the single resonance dominates. We find perfect fit to 
the data with help of eqn.(\ref{disp1}) using one fit-parameter, the resonance 
mass: $\rho(s) \equiv \delta(s-m_{res}^2)$.

In order to see the manifestation of the short-distance phenomena,
we have generated Monte-Carlo configurations on a small (unblocked)
$8^3\times 16$ lattice at $\beta=6$, corresponding to $a^{-1} \sim 2 GeV$.
We use quenched propagators of the clover- and tadpole-improved Wilson fermionic action.
We find that the long-distance part of the correlator (higher moments) is
still well fitted with a single resonance. We reproduce the masses of the 
pseudoscalar ($\eta_c$), vector
($J/\psi$) and scalar ($\chi_{co}$) mesons of the charmonium spectrum with
quite reasonable accuracy.  For lower moments ($\sim$ short-distances) though, 
Monte-Carlo data deviate from the single-resonance curve. 
However, that deviation  proves to be described very
well by the contribution of the excited states - a smooth hadronic continuum
spectrum, which we incorporate via the dispersion relation (\ref{disp1}).

In section 3 we study the lower moments of the correlator, corresponding to the
short-distance physics. They must be reliably reproduced in perturbative QCD,
which means that $one$-loop of free quarks gives the dominant contribution.
The mass of those quarks is the renormalized heavy-quark mass by construction.
Since there is a reliable loop expansion for the lower moments of the correlator
of heavy-quark currents one can study systematic deviations of the coefficients 
of that expansion order by order.

First, we fit the Monte-Carlo data for the lower moments by the 
correlator composed of free Wilson quark propagators. Those propagators 
correspond to the quarks of mass $m_c\,=\,1.37 GeV$ in the continuum theory.
 To be more accurate we study a relation 
between the mass of the free Wilson action and the corresponding fermion mass
of the continuum theory in the finite volume because the Monte-Carlo is done in
the finite (and relatively small) volume. Again, we find evidence that the 
dispersion relation (\ref{disp1}) works well, this time - for the continuum
spectrum of quark-antiquark pair. To specify the
subtraction scheme of the renormalized quark mass we take into account the
two-loop $\alpha_s$-correction. With this refined treatment we obtain in the
minimal subtraction scheme $m_c^{\bar{MS}}(m_c)\,=\,1.22(2) GeV$ from the
vector channel and $m_c^{\bar{MS}}(m_c)\,=\,1.18(2) GeV$ from the pseudoscalar
channel. We attribute the difference in these two values to the absence of
scaling on the given lattice.

\section{Correlator of the heavy-quark currents}

\subsection{Dispersion theory of charmonium}
Consider the correlator of vector currents of charmed quarks:
\begin{equation}
\Pi(q^2)_{\mu\nu}\;\;=\;\;i\int dx e^{iqx}\, <0| T\left\{j_{\mu}(x) j_{\nu}
(0) \right\}|0>  \label{p(q)}
\end{equation}
where $j_{\mu} \,=\, \bar{c}\,\Gamma\,c$.
The correlator (\ref{p(q)}) in the continuum theory satisfies the
standard dispersion relation:
\begin{equation}
\Pi (q^{2})_{\mu\nu}\;\;  = \; \;(q_{\mu}q_{\nu}-g_{\mu\nu}q^2)\,q^2\,
\int ds\; \frac{\rho(s)}{s^2(s\;+\;q^2)}\;+\;d_1 g_{\mu\nu} \;+\;
d_2 (q_{\mu}q_{\nu}-g_{\mu\nu}q^2)      \label{disp_cont}
\end{equation}
The correlator $\Pi (q^{2})_{\mu,\nu}$ is defined up to the polynomial, which
coefficients $d_{1,2}$ have ultraviolet singularities. The polynomial with the 
coefficients $d_1$ and $d_2$ is a local part of the correlator. 
In the $X$-space one has:
\bb
\Pi(x)_{\mu\nu}^{local}\;\;=\;\;d_1\,g_{\mu,\nu}\,\delta^{(4)}(x)\;+\;d_2\,
\left( \partial_{\mu}\partial_{\nu}\,-\,g_{\mu\nu}\partial^2\right)
\,\delta^{(4)}(x)
\ee

$j_{\mu}$ is essentially electromagnetic current of charmed quarks. 
The correlator (\ref{p(q)}) describes electromagnetic vacuum polarization due 
to strong interactions.
The coefficient $d_1$ renormalizes photon mass, while the coefficient $d_2$ 
renormalizes photon wave function. If the ultraviolet singularities are
subtracted on the photon mass shell $q^2=0$ one has $d_1=d_2=0$ for the
renormalized correlator.

The spectral density $\rho(s)$ is proportional to the inclusive cross-section
of the production of charmed and anticharmed hadrons in
$e^{+}e^{-}$-annihilation via the electromagnetic current of heavy quarks. From
experiment one knows that $\rho(s)$ has a large peak due to
$J/\psi$-meson. Higher radial excitations with smaller contribution to the
spectral density overlap with the continuum spectrum of multiparticle states.
The simple and phenomenologically very successful ansatz for $\rho(s)$:
\begin{equation}
\rho_{phen}(s)\;=\;s\,\left(\,f m_{res}^2 \delta(s-m_{res}^2)\,+\,\frac{1}{4 \pi^2}
\theta (s - s_o) \,\right)  \label{modelrho}
\end{equation}
contains a single low-lying resonance of mass $m_{res}=m_{J/\psi}$,
residue $f$, proportional to the electrwomagnetic width of the
$J/\psi$-meson and a smooth continuum spectrum with some effective
threshold $s_o > m_{res}^2$. The coefficient $1/4\pi^2$ comes from one loop 
of free quarks. The fact that continuum hadronic spectrum at high energies 
$s>s_o$ is identical to the perturbative spectrum of free quarks (with small 
radiative corrections) is usually referred to as {\em global duality} \cite{ckt}.
This property is a consequence of asymptotic freedom.

Since the poles and discontinuity of the correlator $\Pi (q^{2})_{\mu\nu}$
are located away from the origin $q^2=0$ the correlator is analytic at $q^2=0$.
Following \cite{svz1} define moments of the polarization operator (\ref{p(q)}) as:
\begin{equation}
{\cal M}_{n}\;\,=\;\,\frac{1}{n!}\;\left\{\left(\,-\,\frac{d}{dq^{2}}\right)^{n} \,
\Pi_{\mu\mu} (q^2) \right\}_{q^{2}=0}  \label{moment}
\end{equation}
From the spectral representation (\ref{disp_cont}) one obtains:
\bb
{\cal M}_{n}\;\,=\;\,-3\;\int ds\; \frac{\rho(s)}{s^{n+1}}  \label{mom_disp}
\ee
Obviously the first two moments ${\cal M}_{0,1}$ suffer from ultraviolet singularities
wheras all the other moments are physical quantities: the ultraviolet singularities
in them can be obsorbed via the renormalization of the Lagrangian parameters,
the strong coupling constant and the charmed quark mass.

The quantity we study below is the ratio $r_n$ of the neighboring moments:
$\,r_{n}\;=\;{\cal M}_{n+1}/{\cal M}_{n} \,$. The phenomenological spectrum 
(\ref{modelrho}) has the following ratios
\begin{eqnarray}
r_n \;&=& \;\frac{1}{m_{res}^2}\;\frac{\eta_n}{ \eta_{n-1}} , \\ \nonumber
\eta_n\;&\equiv&\;1\;+\;\frac{1}{4 \pi^2 n f}\,
\left(\frac{m_{res}^2}{s_o}\right)^n                         \label{modelmom}
\end{eqnarray}
One can see that the continuum contribution to the higher moments rapidly 
decreases. Higher moments are saturated by large distances, where the 
single low-lying resonance dominates. The ratios (\ref{modelmom}) corresponding
to the single-resonance approximation to the correlator (\ref{p(q)}) are
particularly simple:
\begin{equation}
r_{n}^{res}\;\;=\;\;\frac{1}{m_{res}^{2}}   \label{r_res}
\end{equation}
where $m_{res}$ stands for the resonance mass. This is illustrated in Fig.1.

Lower moments receive substantial contribution from the continuum spectrum.
They probe the vicinity of the origin $q^2=0$. From the QCD point of view that
region is in the domain of asymptotic freedom: the nearest threshold due to a 
pair of charmed quark and antiquark is away by $4 m_{c}^{2} \gg \Lambda_{QCD}^{2}$.

The function $\Pi (q^{2})$
is reliably calculable in perturbative QCD near $q^2=0$. 
Low-energy thresholds due to multi-gluon intermediate states, which
appear on the $4$-loop level, are known to give small contributions in the
case of heavy quarks \cite{svz1}. They may be eliminated altogether by giving
different masses to the two quarks of the current.
$\Pi (q^{2}=0)$ is a function of the renormalized
heavy quark mass $m_c(\mu)$ and $\alpha_{s}(\mu)$. The relevant
renormalization point here, $\mu \,= \,2 m_c$, is high enough to ensure the
validity of the QCD perturbation theory: $\alpha_{s} (4 m_{c}^{2})\,\simeq\,0.2$.
Then the applicability of perturbative QCD near $q^2 = 0$ implies the
following expansion for the ratios $r_n$:
\begin{equation}
r_{n}\;=\; \frac{1}{4 m_{c}^{2}} \,\left( a_n\;+\; b_n\, \alpha_{s}(4 m_{c}^{2}) 
\;-\; c_n \, \frac{G^{(2)}}{\left(4 m_{c}^{2}\right)^2} \right) \label{rqcd}
\end{equation}
where $\{a_n, b_n, c_n \}$ are known numbers \cite{svz1}.  
The term $\sim a_n$ comes from one loop of free charmed quarks.
The term $\sim b_n$ comes from the two-loop diagrams corresponding to
one-gluon exchange. The gluon condensate 
$G^{(2)}\,\equiv\,<0|(\alpha_{s}/\pi)\,G_{\mu\nu}^{a}G_{\mu\nu}^{a}|0>$ is 
the vacuum average of the first nontrivial operator in the Wilson 
operator-product expansion for the correlator of heavy-quark currents.
The coefficient $c_n$ originating from the Wilson coefficient function
starts with one loop of the heavy-quark propagators.

The lower ratios $r_{2,3,4}$, originating from short distances, are well
reproduced by QCD perturbation theory. The typical virtuality corresponding to the $n$th moment is $\sim \, 4m_{c}^{2}/n$. The coefficient $b_n$ grows
with $n$. At large $n>8$ the perturbation-theory based expansion (\ref{rqcd}) 
is irrelevant. The gluon condensate shows up in the intermediate ratios 
$r_{5,6,7}$ \cite{svz1}, sensitive to larger distances and,
hence, nonperturbative fluctuations. It represents the first term of the operator product expansion. 
The operator-product expansion thus describes
a crossover from the short-distance region to the large-distance
region, in the vicinity of the domain of asymptotic freedom. The contribution of 
the gluon condensate into the physical quantity $r_n$ is renormalization-scale 
independent. To extract the value of the gluon condensate from $r_n$ one has
to know the parameters of perturbative QCD. The better one knows $\alpha_s$ and
$m_c$ from the lower moments $r_{2,3,4}$ - the higher the accuracy that can be 
reached in the calculation of the gluon condensate from $r_{5,6,7}$. 

The phenomenological parameters of hadronic spectrum (\ref{modelrho}), based 
on experimental data on the inclusive cross-section of charm-anticharm 
production in $e^{+}$$e^{-}$-annihilation, are :
\bb
4\pi^2 f\;\simeq\; 0.6\;,\;\;\;\;\;m_{res}\;\simeq\;3.1\;GeV\,,\;\;\;\;\;
s_o\;\simeq\;\left(4\;GeV \right)^2          \label{spectrum}
\ee
The moment ratios of the corresponding correlator are shown on FIg. 1.
One can see that the higher moments $n \geq 7$ of the complete correlator,
shown by bursts, concide with the single-resonance approximation to that
correlator - the contribution of the $J/\psi$-meson, shown by squares.
The lower moments, corresponding to the short-distance part of the correlator,
 deviate from the single-resonance line. They are well reproduced with one 
loop of free quarks of mass $m_c = 1.26 GeV$, which moment ratios are shown by
diamonds. This way the mass of charmed quark was first estimated 
within QCD in \cite{svz1}. 

In order to make this estimate more accurate one
has to know the contribution of continuum hadronic spectrum very well,
since it is large in the lower moments. However,
this is demanding with respect to experimental data on the inclusive
cross-section of charm-anticharm production in the $e^{+}e^{-}$-annihilation.
Additional uncertainty here is due to the fact that at high enrgies one has 
to separate the bigger part of that inclusive cross-section, coming from the
electromagnetic current of heavy quarks, from the smaller part of that 
cross-section, coming from the electromagnetic current of light quarks. 
In order to avoid the 
uncertainty associated with the hadronic continuum alternative ways to 
determine charmed quark mass have been developed. They are based on the 
nonrelativistic approximation to QCD. Potential models \cite{ynd} are particularly
successful for $b$-quarks. Charmed quarks are light enough to exhibit substantial
relativistic effects. In this work we choose to stay within the original QCD rather then to
study nonrelativistic approximaitons to it. In order to avoid the uncertainty 
due to hadronic contiuum we evaluate the correlator (\ref{p(q)}) by Monte-Carlo
in lattice theory. The lattice calculation requires much less experimental data.
One only has to know the mass of the low-lying resonance in order to fix the 
scale of the lattice theory.

The fact that the lower moments are well reproduced in perturbative QCD implies
an integral relation, called {\em sum rule}, between the parameters of 
perturbative QCD and phenomenological parameters of hadronic spectrum:
\bb
\int \frac{ds}{s^{n+1}}\; \rho_{QCD}(s)\;\;=\;\;\int \frac{ds}{s^{n+1}}\; 
\rho_{phen}(s)        \label{sumrule}
\ee
where $n = \{2,3,4\}$ and $\rho_{QCD}(s)$ is the spectal density corresponding 
to the multiloop expansion of perturbative QCD (\ref{rqcd}). The parameters of 
hadronic spectrum, consistent with the sum rule (\ref{sumrule}) and 
experimental data (available in the case of vector channel),
are those in eqn.(\ref{spectrum}).

One important implication of the sum rules (\ref{sumrule}) is that smooth 
hadronic continuum spectrum, which form coincides with the corresponding 
perturbative expression, {\em must} be incorporated in the model of the 
hadronic spectral density. At high energies the continuum spectrum dominates 
over the contribution of few excited states. The high-energy asymptotics of
the hadronic spectrum must be the same as predicted by the perturbative QCD.
Therefore, a finite number of narrow resonances is ruled out as a model
of hadronic spectral density of the correlator (\ref{p(q)}). 

One should emphasize that light quarks do not participate in the derivation of 
the spectrum (\ref{spectrum}) in a sense that the expansion (\ref{rqcd}) does not
contain diagrams with light quarks. Incorporation of light quarks
would bring more accuracy, in general, but the contribution of light quarks 
into the lower moments is essentially small, as they appear in higher orders of 
the reliable perturbative expansion in powers of $\alpha_s(4m_{c}^{2})$ \cite{svz1}. 

\subsection{Dispersion relation on the lattice}
In the previous section we have pointed out that the short-distance part of 
the correlator ($r_{2,3,4}$) receives substantial contributions from the high-energy 
(continuum) part of the spectrum. When analysing the data on the correlator
in lattice QCD we would like to know how the correlator with a given
spectrum looks like in the lattice theory. How the finite volume and finite
cutoff modify the function $\Pi (x)$ with the spectral density as in
(\ref{modelrho})? 

We explore the following naive extension of the 
dispersion relation (\ref{disp_cont}) to the lattice theory:
\begin{equation}
\Pi (q_{\mu}, a,L)\;\;  = \; \;\int ds\; \frac{\rho(s, a, L)}{s\;+\;
\frac{4}{a^2} \,\sum_{\mu}^{4} \sin^{2}(q_{\mu} a/2)}  \label{disp_lat}
\end{equation}
In principle, one should expect the spectral density in (\ref{disp_lat}) to
have lattice artifacts. In practice we find that the simple modification of
the denominator as in (\ref{disp_lat}) accounts for the major part of
systematic deviations due to finite $L$ and $a$. Eqn.(\ref{disp_lat})
allows one to fit Monte-Carlo data by a complicated spectrum with several
resonances and a smooth continuum.

In the case of a single resonance $\rho(s) = \delta (s - m^{2}_{res})$ 
eqn.(\ref{disp_lat}) becomes a familiar expression for the lattice propagator 
of a free field:
\begin{equation}
\Pi^{res} (q^{2}, {\bf n})\;\;  = \; \;\frac{1}{m_{res}^2\;+\;
\frac{4}{a^2} \,\sum_{\mu}^{4} \sin^{2}(q_{\mu} a/2)}  \label{p(q,n)}
\end{equation}
where we have introduced a four-vector ${\bf n}$ of unit lengh: $q_{\nu}\equiv
q n_{\nu}$, so that $\sum_{\nu=1}^{4} n_{\nu}^{2}=1$. The extra dependence of the
correlator on $n_{\nu}$ is an ${\cal O}(a^2)$-effect.

To see how well eqn.(\ref{disp_lat}) works for the continuum spectrum we 
fit in section 3 the lattice correlator in the free Wilson quarks approximation 
(which has contuum spectrum) to the eqn.(\ref{disp_lat}) with the quark-antiquark 
spectral density of the continuum theory. It proves to work well on finite 
lattices of different size.

Since we are interested in the short-distance behavior of the correlator we
have to address the question of subtractions. At short distances the correlator
is singular. In the continuum theory the ultraviolet singularities are 
proportional to the polynomial of finite order in the momentum space. The finite
order of that polynomial is a consequence of the renormalizability of the 
theory. The renormalized correlator is the one with the infinite parts of the
local terms subtracted. In the continuum theory the difference between subtracted 
and unsubtracted correlator is polynomaial in $q^2$. For example, in the 
correlator of the vector currents (\ref{p(q)}) that polynomial is quadratic:
$d_1\,+\,d_2 q^2$. The factor $q^2$ here is essentially the inverse Laplacian of
the continuum theory. The subtraction polynomial contributes only to the first
two moments. The other moments do not receive any contribution at all from the 
subtraction polynomial. 

Let us see what the difference is between the subtracted and unsubtracted 
lattice dispersion relation (\ref{disp_lat}). The subtraction:
\begin{equation}
\frac{1}{s\;+\;\frac{4}{a^2}\sum_{\mu}^{4} \sin^{2}(q_{\mu} a/2)}\;-\;
\frac{1}{s}\;\;=\;\;-\;\frac{\frac{4}{a^2}\sum_{\mu}^{4}\sin^{2}(q_{\mu}a/2)}
{s\left(\,s\;+\;\frac{4}{a^2}\sum_{\mu}^{4} \sin^{2}(q_{\mu} a/2)\,\right)}
\end{equation}
generates non-polynomial terms $q^{2}_{lat}\,\equiv \,\frac{4}{a^2}\sum_{\mu}^{4}
\sin^{2}(q_{\mu} a/2)$ instead of $q^2$ in the numerator. To perfom a subtraction
in the lattice theory one would have to decompose the correlator (\ref{disp_lat})
in the following form:
\begin{eqnarray}
\Pi (q^{2})\;\;&=&\;\;\tilde{d}_1\;+\;\tilde{d}_2 \,\frac{4}{a^2}\sum_{\mu}^{4} 
\sin^{2} (q_{\mu} a/2)\;  \nonumber  \\
&+&\;\left(\frac{4}{a^2}\sum_{\mu}^{4} \sin^{2}(q_{\mu}
a/2)\right)^{4} \int ds\; \frac{\rho(s)}{s^2\;(s\;+\;\frac{4}{a^2}
\sum_{\mu}^{4} \sin^{2}(q_{\mu} a/2))}                             \label{cs}
\end{eqnarray}
The factor $q^{2}_{lat}$ is the inverse of the standard lattice Laplacian:
\bb
\int \frac{d^{4}q}{(2 \pi^4)}\;e^{iqx}\;\frac{4}{a^2}\sum_{\mu}^{4} 
\sin^{2} (q_{\mu} a/2)\;\,=\;\,\left(\,2 \delta(x)\,-\,\delta(x-a)\,-\,
\delta(x+a) \, \right)\,/\,a^2             \label{subtr_x}
\ee
Therefore the coefficient $\tilde{d}_2$ in (\ref{cs}) is precisely the 
renormalization of the photon wave function in the case of vector currents of 
charmed quarks. The renormalized correlator would be given by eqn.(\ref{cs}) with
$\tilde{d}_1\,=\,\tilde{d}_2\,=\,0$, considered as a function of the renormalized
Lagrangian parameters.

The deviation of the $q^{2}_{lat}$-term
from a polynomial $q^2$ is an ${\cal O}(a^2)$-cutoff effect. 
It generates ${\cal O}(a^n)$ systematic deviations
in the moments, which decrease rather fast with $n$. For a single resonance, 
for example, in the $1$-dimensional case one can find analytically explicitly:
\begin{equation}
{\cal M}_{n}^{res}\;\;=\;\;\left\{\;1\;+\;
\frac{n!}{(2 n)!}\,(a m_{res})^n\;\right\}\;\frac{1}{(a m_{res})^{2n}}
\end{equation}
In the next sections we will see the size of the contribution of the subtraction 
term $q^{2}_{lat}$ into the moments on finite lattices. The ratios $r_n$ for 
$n\geq 3$ prove to be practically not affected by subtractions.

\section{Moments on a finite lattice}
\subsection{Single resonance}
The definition of the moments (\ref{moment}) is designed in such a way
that the ratios $r_n$ of the neighboring moments of the correlator,
saturated by a single resonance, are all equal to the inverse mass squared
of that resonance. This suggests a simple way to determine the resonance
mass numerically from given data on the correlator (\ref{p(q)}). The
property (\ref{r_res}) however, holds in the continuum theory, that is
in the limit $a \rightarrow 0 $ and $L \rightarrow \infty$. One has to see 
how much the moments of a single resonance are distorted by the finite $L$ 
and $a$.

The extension of the definition of moments (\ref{moment}) to the case of a
finite lattice (with broken Lorentz invariance) was suggested in \cite{b}:
\begin{equation}
{\cal M}_{n}\;=\; \frac{1}{2^{2n}\, n! (n+1)!}\,
\int dx^4 \, x^{2n} \,\Pi(x)  \label{xmoment}
\end{equation}
In the continuum theory eqns. (\ref{moment}) and (\ref{xmoment}) are
identical. An alternative to (\ref{xmoment}) would be to project the
correlator (\ref{p(q)}) to $\vec{q}=0$ and define the moments as derivatives
with respect to $q^o$. In this work we explore the Lorentz invariant
definition (\ref{xmoment}).

We have evaluated the moments (\ref{xmoment}) of a single resonance
(\ref{p(q,n)}) numerically on a given lattice.
Figs. 2,3,4 show the moments of a single resonance for different volumes.
One can see very strong finite volume effects. For small values of $L$ and
$a m_{res}$ the moments are far away from the straight line (\ref{r_res})
of the continuum theory, illustrated by squares in Fig.1. As one increases
the volume, the moments approach a straight line corresponding to
(\ref{r_res}).
This works better for larger values of $a m_{res}$ (Fig.4), corresponding
to more compact resonance-objects. The constant behavior of $r_n$ is clearly
seen for $m_{res}=1.5$ in the whole range of the ratios of interest
$r_{1\div 7}$ on a large lattice $32^4$. For larger $n$ the moments
deviate from the prediction of the continuum theory (\ref{r_res}) because
higher moments probe larger distances, hence they are more sensitive to the
boundary of the box. Figure 5 shows the behavior of the moments for different
resonance masses $m_{res}=0.5\div 2$ on the lattice $16^3 * 32$. Again, one can
see that finite-volume effects disappear for larger values of the resonance
mass in the units of the lattice spacing.
It is important to realize that these finite-size effects do not represent
any loss of information. The curves are universal and may (will) be used to
fit Monte-Carlo data on the same lattices.

Figs. 2,3,4 show the moments of the subtracted correlator as well. The subtracted 
correlator has vanishing first two moments ${\cal M}_{0,1}=0$. The ratio 
$r_1 \,=\,{\cal M}_{2}/{\cal M}_{1}$ of the subtracted correlator is not to be 
considered. The two solid lines on Figs. 2, 3, 4, corresponding to the subtracted
and unsubtracted correlators, are indistinguishable except for the second 
ratio $r_2$. The line corresponding to the subtracted correlator is always the
one that terminates at $n=2$. In the continuum theory the ratio $r_2$ is not
affected by subtractions because the subtraction polynomial is $O(q^2)$. On the
lattice the subtraction term $q^{2}_{lat}$ gives contributions to higher 
moments. Those contributions are noticable in the ratio $r_2$ and negligable 
in higher ratios. The deviation between the subtracted and
unsubtracted curve in the ratio $r_2$ is order $O(a^2)$ effect. 
One can see that the deviation in the ratio $r_2$ decreases
with the resonance mass, because the decrease in the resonance mass here 
essentially implies the decrease in the lattice spacing $a$. In the limit 
$a \rightarrow 0$ one has $q^{2}_{lat} \rightarrow q^2$, hence the ratio $r_2$
becomes insensitive to subtractions. 

Concentrating on the cutoff effects we extract the resonance mass from the 
ratio $r^{\ast} \equiv r_n (L\rightarrow \infty)$, observed on the biggest 
available lattice $32^4$, by looking for a plateau like the one on the upper 
curve of Fig. 4 for $n<8$. Increasing the volume would extend this plateau to 
larger $n$, while its position would not change. This would be our procedure to 
fit the resonance mass $\bar{m}_{res}\equiv (r^{\ast})^{-1/2}$ via the
analysis of the moments. However, even if the lattice is large enough to 
exhibit a perfect plateau for $r_n = r^{\ast}$,
corresponding to the continuum theory, the position of that plateau gives
$1/m_{res}^{2}$ only approximately because of finite-$a$ effects, as one
can notice by inspecting the upper lines of Figs. 3 and 4.
The resonance masses $\bar{m}_{res}$ are plotted on Fig. 6 versus the input
mass $m_{res}$. The fit of that data gives the following
${\cal O}(a^2)$-systematic deviation\footnote{
All the dimensional quantities in the lattice theory are in units of the
lattice spacing $a$ throughout this paper, although we show the factor $a$
explicitly sometimes to draw attention to the dependence on it.}:
\begin{equation}
\frac{1}{\bar{m}_{res}^{2}}\;\;=\;\;\frac{1\;+\;0.04\, m_{res}^{2}}
{m_{res}^{2}}  \label{Oa2_res}
\end{equation}
The fluctuations of the data Fig. 6 around the fit (\ref{Oa2_res}) near
$m_{res} = 1$ are caused by our uncertainty in determining
the plateau $r_n = r^{\ast}$, which is not
well-established yet on a $32^4$-lattice for those values of $m_{res}$.
The value $m_{res} = 3$ is obviously
too large to be fitted by the ${\cal O}(a^2)$-law (\ref{Oa2_res}).
The systematic correction (\ref{Oa2_res}) is rather small for $m_{res}\leq 1$,
so the present Monte-Carlo data for light quarks (where the resonance masses
in units of the lattice spacing are typically less than $1$) does not see it.

In Fig. 7 we compare the resonance mass $\bar{m}_{res}$ with the conventional
way of extracting the resonance mass from the correlator (\ref{p(q)}) -
a fit to $cosh(t-L/2)$ at vanishing three-dimensional momentum $\vec{q}=0$.
One can see that the resonance mass $\bar{m}_{res}$, extracted from the
moments is always a better approximation to the expected result
$m_{res}$. 

\subsection{Monte-Carlo data}
In this section we analyze the moments of the correlator (\ref{p(q)}),
computed by Monte-Carlo in the quenched approximation, first on a
large but coarse lattice obtained by blocking, then on a small but fine
lattice. These two studies allow us to check separately our numerical
control of large and small moments.

The fermionic action we use $S_F\,=\,S_W\,+\,S_{SW}$ has the Wilson term $S_W$
and the clover term $S_{SW}$ \cite{Clover}:
\beqa
&&S_W=\sum_{x}\;\{\;\bar{\psi}(x)\,\psi(x) \nonumber \\
&-&\kappa_{W}\,\sum_{\mu}\,\left[\,\bar{\psi}(x)(r-\gamma_{\mu})
U_{\mu}(x) \psi(x+\hat{\mu})\,+\,\bar{\psi}(x+\hat{\mu})(r+\gamma_{\mu})
U_{\mu}^{\dagger}(x) \psi(x) \,\right]\,\} \label{S_W}
\eeqa
\beqa
S_{SW}\;\;=\;\;-\;\frac{r\,\kappa_{W}}{2}\;c\;\sum_{x,\mu,\nu}\;
\bar{\psi}(x)\;F_{\mu \nu}(x)\,\sigma_{\mu \nu}\;\psi(x)     \label{S_SW}
\eeqa
where $F_{\mu \nu}$ is a field-strengh tensor as the following sum of four
plaquettes in the $\mu \nu$ plane around point $x$:
\bb
F_{\mu \nu}(x)\;\;=\;\;\frac{1}{8}\;\sum^{4}_{\Box=1}\;\left[\;
U_{\Box \mu \nu}(x)\;-\;U^{\dagger}_{\Box \mu \nu}(x)\;\right]
\ee
The action $S_F$ does not have terms of order $O(a)$ in the expansion in
powers $a$ as $a \rightarrow 0$ for $c = 1$ \cite{Clover}. The Green 
functions computed with the help of action $S_F(c = 1)$ do not differ 
from their limit $a \rightarrow 0$ in the order $O(a)$ provided the 
fermionic fields $\psi(x)$ are related to their $a\rightarrow 0$ limit
as follows:
\bb
\psi\;\;\rightarrow\;\;\left( \,1\;-\;\frac{r a}{2}\,\left(z \gamma_{\mu}
D_{\mu}\,-\,(1-z) m^w \right)\, \right) \psi   \label{rot}
\ee
where $m^w$ is the bare quark mass, parameter of the Wilson propagator, and
 $D_{\mu}$ is the lattice covariant derivative and $z$ arbitrary 
parameter, corresponding to the possibility of using the equations of motion 
 \cite{MSV}. We choose $z=0$.

However, the systematic deviations of the order $O(g^{2n} a)$ are generated 
by the action $S_F$. With the hope to diminish them we would like to 
eliminate tadpole diagrams within the mean-field approximation \cite{lm}. 
The tadpole diagrams make average links differ from one: $<U_{\mu}> \neq 1$.
The suggestion of \cite{lm} is to absorb these factors by the nonperturbative
renormalization of the parameters of the lattice action. Identifying 
$<U_{\mu}>\,=\,<U_{\Box}>^{1/4}$ in the mean-field approximation, we devide
every link in (\ref{S_W}), (\ref{S_SW}) by $<U_{\Box}>^{1/4}$. This implies 
that we use a new $\kappa = \kappa_W /<U_{\Box}>^{1/4}$ in $S_F$ and 
nontrivial value of the coefficient $c$ against clover term (\ref{S_SW}): 
$c\,=\,<U_{\Box}>^{-3/4}$. 

\subsection{Blocked lattice}
We have obtained a set of blocked $SU(3)$ configurations from the QCD-TARO
collaboration \cite{jap}. Starting from a $32^3\times 64$ lattice at
$\beta=6$, corresponding to $a^{-1}\simeq 2 GeV$ \cite{bs}, two blocking
steps were performed. This brought the lattice size to $8^3\times 16$, with
a rather large lattice spacing $a^{-1} \sim 500 MeV$.
The blocking process preserves information about large distances, but
of course does not allow us to study short-distance effects.
In \cite{bdef} the
clover- \cite{Clover}-and-tadpole-improved \cite{lm} quark propagators were
computed on these blocked configurations. These propagators
were used in \cite{b} to evaluate the moments of the correlator of
vector currents.  Here we fit those moments by the moments of a single
resonance computed in the same box in accordance with section 1.1.

Fig. 8 shows the moments of the correlator of pseudoscalar currents.
Three sets of data correspond to three different values of $\kappa =
\{ 0.1111, 0.1031, 0.0983 \}$. Statistical errors are shown on the graph.
The upper curve, corresponding to relatively light quarks $\kappa=0.1111$,
($m_{pion}\sim 500 MeV$) comes from 28 configurations. The other two sets
of data correspond to heavy quarks, where statistical errors become
small for already $\sim10$ configurations. The solid lines connect
the ratios of the moments of a single resonance, computed on the same
lattice size $8^3\times16$. The input mass of that resonance $m_{res}$ is 
a fit parameter, which we find to be: $m_{res}\,=\,\{1.08, 1.95, 2.36\}$.
In this approach, the fitted resonance mass automatically
corrects for the systematic deviation (\ref{Oa2_res}).
 One can see that the solid lines fit the Monte-Carlo data
perfectly at all points despite the nontrivial shape of the curves.
(The first two ratios $r_{1,2}$ are not to be considered
because they are subject to subtractions as explained in the previous
section.)  This implies that the Monte-Carlo data on the correlator is
saturated entirely by a single resonance, which is of no surprise:
the blocked lattice sees large distances only.

\subsection{Fine lattice}
In order to be sensitive to the high-energy part of spectrum, which
controls the behavior of the correlator at short distances, we have
generated $SU(3)$ Monte-Carlo configurations on a small
$8^3*16$-lattice at $\beta=6$ ($a^{-1}\simeq 2 GeV$). Quark
propagators were computed for the
clover-and-tadpole-improved action.

Fig.9 shows Monte-Carlo data from $20$ configurations for the moments of
the pseudoscalar (squares), vector (octagons) and scalar (diamonds)
correlators for $\kappa=0.1000$.
Statistical errors (shown in Fig. 9, 10, 11) are rather small because the 
quarks are heavy. $3000$ sweeps were used for thermolization and $1000$
sweeps separated configurations for decorrelation. No correlations from one
configuration to the next were noticed.

We fit the higher moments $n\geq 8$ by the moments of a single resonance,
computed in an $8^3\times16$ box (solid lines). Fig.9 shows three sets of
double solid lines, corresponding to the three channels: pseudoscalar, vector 
and scalar. The lines are doubled to show the impact of
statistical errors, estimated by jack-knife, on the determination of the
resonance mass. The masses of the resonances in units of the lattice spacing
are shown in the tables.
 The spectrum of the charmonium ground states thus obtained
is shown in Fig.10. It is gratifying to reproduce the charmonium spectrum
with such a reasonable accuracy on a small lattice $8^3*16$. Current
simulations of heavy systems \cite{fermilab} are performed on significantly
larger lattices.
The reason for
this success is that the charm-anticharm bound states are small objects:
the size of the $J/\psi$-meson is $\sim 600MeV$. Also the splitting between
$S$-wave states and $P$-wave states has a perturbative nature within
the nonrelativistic approximation, and is accounted for by the leading
loop of free quarks.

The new feature that one observes in Figs. 9, 11 is that the lower moments
deviate from the single-resonance curve. This was not seen at all on the
blocked lattice. We attribute this fact to the short-distance contributions.

We have modelled the hadronic spectral density as in (\ref{modelrho}) with 
the phenomenological parameters of the charmonium spectral density 
(\ref{spectrum}) and computed the moments of the corresponding correlator
on an $8^3\times16$ lattice in accordance with (\ref{disp_lat}).
The phenomenologically obtained lattice moments, shown in Fig. 11 by the
dashed line, fit the Monte-Carlo data perfectly. It is obviously the contribution
of the high-energy part of the hadronic spectrum which accounts for the
deviation of the Monte-Carlo data from the single-resonance curve at
lower moments. The contribution of the continuum spectrum in the smaller
ratios is obviously quite large, as it accounts for the strong deviation of 
the Monte-Carlo data from the single-resonance curve. In contrast to that, 
the continuum contribution is quite small for $n>6$ as compared to the 
resonance one. The helps to determine the resonance mass with better accuracy.

Fig. 12 compares the two ways of extracting the resonance mass from 
Monte-Carlo data. The upper curve shows fit to the moment ratios with the 
single-resonance appoximation to the correlator. The curve flattens at higher 
moments corresponding to large distances, where the single resonance dominates.
The conventional fit of the zero-momentum component of the correlator to 
$cosh[(t-L/2) m_{res}]$ is shown on Fig. 13.
We do not see a wide horizontal plateau at large $n$ on Fig. 12 (or at 
large $t$ on Fig. 13) because the volume of the box $8^3 \ast 16$ is small and
the contribution of the high-energy part of the spectrum is noticeable.
The bottom curve on Fig. 12 takes into account the contribution of the high-energy part
of the spectrum. We use ansatz (\ref{modelrho}) in the dispersion relation 
({\ref{disp_lat}}) with fixed residue of the resonance and the ratio of the 
effective continuum threshold to the resonance mass. Those two parameters are
taken to have phenomenological values as in (\ref{spectrum}). The resonance mass
then remains a fit parameter. The procedure is applied to the pseudoscalar and
scalar channels with similar values of the phenomenological dimensionless 
spectrum parameters $f$ and $s_o / m^{2}_{res}$. Our typical statistical
error in the vector and pseudoscalar resonance masses is $\sim \pm 1\%$, while
in the scalar channel it is $\sim \pm 5\%$.
The values of the resonance mass obtained this way are shown in the tables. 
The continuum contribution helps to stabilize the resonance mass at smaller $n$.
On bigger lattices one can probe longer distances, where the continuum contribution
is very small, and obtain stable curves for the resonance mass in a wide range 
of $n$ without using the continuum contribution. 

The model of the spectrum at high energies in the form of a smooth perturbative
continuum is, in principle, better than the two-resonances fit because this
model respects the global duality, which requires that the spectral density
does not decrease at high energies. In the world with no light quarks 
(corresponding to the quenched simulations) the smooth continuum models 
an infinite (in the limit of infinite cutoff) series of resonances, while 
in the world with light quarks the effective continuum threshold is close
to the $D$-$\bar{D}$-meson pair production in the charmonium vector channel.
Since the model of the smooth perturbative continuum is somewhat rough in 
any case, one should not consider {\em bad} quantities, which are very sensitive to the 
precise form of the continuum spectrum. {\em Good} quantities are those 
related to intergrals over the spectral density of the form of 
eqn.(\ref{sumrule}).

The value $\kappa =0.1060$ reproduces $m_{\eta_c}=2.979 GeV$, 
for the lattice spacing $a^{-1} = 1.90 GeV$, and it reproduces 
$m_{J/\psi}=3.097 GeV$ for $a^{-1} = 1.92 GeV$. These values of the lattice 
spacing are consistent with the string-tension measurements at $\beta=6$ \cite{bs}.
 Neighboring values of $\kappa$ are also possible,
 if one changes the lattice spacing accordingly; to leading order the changes
 in $\kappa$ and in $a$ compensate each other, and our physical results on
 the charmonium spectrum and the charmed quark mass are not affected. The
value of the mass splitting within the $S$-wave that we obtain with the 
tadpole improved clover action is $m_{J/\psi} - m_{\eta_c}\, =\, 87\pm 9\,MeV$.

Fig. 11 is analogous to Fig. 1. It demonstrates the important
fact that the high-energy part of the spectrum and the low-energy modes 
can be seen on the same lattice of limited size. In the next section we study
the lower moments. We fit them with the corresponding expressions of perturbative 
QCD.

\section{The renormalized heavy-quark mass}
\subsection{Off-shell mass on the lattice}
While the large-distance behavior ($n \gg 1$) of the correlator (\ref{p(q)}) is
dominated by a single resonance, the short-distance behavior ($n\sim 1$) is
determined by one-loop of heavy quarks. Since the Monte-Carlo simulation is 
done in the lattice theory with noticeable finite-volume and cutoff effects,
we have to fit Monte-Carlo data with the loop-expansion of perturbative QCD
evaluated in the lattice theory as well. In other words, we have to know 
the systematic deviations in the coefficients $a_{n}(a,L)$, $b_{n}(a,L)$, 
$c_{n}(a,L)$ of eqn.(\ref{rqcd}).
As the two-loop $\alpha_s$-correction is small in $r_{2,3,4}$ we first 
concentrate on one loop of free quarks and clarify its systematic deviations 
completely.

Consider the propagator of free Wilson fermions:
\begin{equation}
S(q)\;\;=\;\;\frac{-i\,\sum_{\mu}^{4} \gamma_{\mu}\,sin(q_{\mu})\;+\;
r\,\sum_{\mu}^{4} \left(1 - cos(q_{\mu}) \right)\;+\;m_{c}^{w} }
{\sum_{\mu}^{4} sin^{2}(q_{\mu})\;+\;\left[\,\sum_{\mu}^{4}
r\,\left(1 - cos(q_{\mu}) \right)\,+\,m_{c}^{w}\, \right]^2} \label{fer_prop}
\end{equation}
At the tree level the fermion mass $m_{c}^{w}$ is related to the hopping parameter
$\kappa$ in a simple way:
\begin{equation}
a m_{c}^{w} \;\;=\;\;\frac{1}{2 \kappa}\;-\;4 r   \label{m(k)}
\end{equation}
The parameter $r$ is in the interval $0\leq r\leq 1$. We use the notation 
$m_{c}^{w}$ to emphasize that this mass is a parameter of the lattice theory.

Evaluating numerically the moments of the correlator (\ref{p(q)}) in the
free-fermions approximation (with Wilson propagators (\ref{fer_prop}) for $r=1$),
 we obtain the ratios $r_n$ as shown on Fig. 14
for periodic boundary conditions. Again, as in the case of the resonance,
one observes a strong volume dependence: higher moments, corresponding to
long distances, feel the boundary of the box. As the volume increases
$8^4\rightarrow 32^4$ the function $r_n$ approaches the form of the continuum
theory shown in Fig. 1 by diamonds. In the continuum theory the moment ratios
of vector, pseudoscalar and scalar correlators in the one-loop approximation
are known to be \cite{svz1}:
\begin{eqnarray}
r_{n}^{vec}\;&=&\;\frac{n^2-1}{n^2+3n/2}\,\frac{1}{4 \bar{m}_{c}^{2}}\;,
\label{rvec_cont}\\
r_{n}^{psc}\;&=&\;\frac{n-1}{n+1/2}\,\frac{1}{4 \bar{m}_{c}^{2}}\;, 
\label{rpsc_cont}\\
r_{n}^{sca}\;&=&\;\frac{n-1}{n+3/2}\,\frac{1}{4 \bar{m}_{c}^{2}}
\label{rsca_cont}
\end{eqnarray}
with the universal property:
\begin{equation}
r_{n\rightarrow\infty}\;\;\rightarrow\;\;\frac{1}{4 \bar{m}_{c}^{2}} 
\label{r_f_asym}
\end{equation}
where $\bar{m}_c$ is the tree-level quark mass in the continuum theory.
On Fig. 15 we compare the numerically computed moments of the pseudoscalar
correlator for free Wilson quarks on the $32^4$ lattice for
$\kappa=0.1,\, (m_{c}^{w} = 1)$ (solid line) with the corresponding values
of the continuum theory (\ref{rpsc_cont}) (diamonds) for $\bar{m}_c = 1$.
The discrepancy between the solid line and the diamonds is too large to
remain unexplained. The asymptotics of the solid line at large $n$,
determined by the large-distance ($q\rightarrow 0$) behavior of the correlator,
deviates from (\ref{r_f_asym}). To understand this we explore the fermion
propagator (\ref{fer_prop}) at small momenta. In the limit $q\rightarrow 0$ 
one finds the following behavior of the denominator in (\ref{fer_prop}):
\begin{equation}
\sum_{\mu}^{4} sin^{2}(q_{\mu})\;+\;\left[\,\sum_{\mu}^{4} r\,\left(1-cos(q_{\mu})
\right)\,+\,m_{c}^{w}\, \right]^2\;\;\rightarrow\;\;q^2\;\left(\,1\;+\;r
\cdot m_{c}^{w} \,\right)\;+\;(m_{c}^{w})^{2}  
\end{equation}
One can see here the lattice renormalization of the fermion wave-function
 $\sqrt{1 + r a m_{c}^{w}}$,
reported previously \cite{lm}, as well as the following tree-level lattice
renormalization of the fermion mass:
\begin{equation}
r_{n\rightarrow\infty}\;\;\rightarrow\;\;\frac{1}{4 (a \bar{m}_c )^2}\;\;=\;\;
\frac{1}{4 (a m_{c}^{w} )^2}\;\left(\,1\;+\;r\, a\, m_{c}^{w} \, \right)
\label{m(k)cont}
\end{equation}
Eqn. (\ref{m(k)cont}) is ${\cal O}(a)$ systematic deviation. The fermion
mass of the continuum theory is related to $\kappa$ in a way more complicated
than (\ref{m(k)}) even at the tree level. The ratios $r_n$ of the continuum
theory corresponding to $\bar{m}_c (m_{c}^{w}=1)$ are shown in Fig. 15 by the 
dashed line. Deviation of the solid line from the dashed line at $n\sim 10$ is
a finite-size effect, which would disappear on lattices larger than $32^4$.
The equality of the asymptotics ($n\rightarrow \infty$) of the solid line and
the dashed line is expected by construction. The remarkable overlap between
the solid
and dashed lines over a wide range of $n$ is less trivial. It means that the
correlator (\ref{p(q)}) composed of free Wilson fermions behaves as in
the continuum theory over a wide range of distances, provided the
relation (\ref{m(k)cont}) between the masses $\bar{m}_c$ and $m_{c}^{w}$ is used.
Fig. 16 illustrates how well this works for different $\kappa$. 

The tree-level relation (\ref{m(k)cont}) is exact to all orders in $a^n$. This 
is because the mass $\bar{m}_c$ is the {\em off-shell} mass, defined at $q^2=0$
in contrast to the pole mass. As promised in the introduction, we never use
the pole mass. 

Figs. 14 and 16 show the moment ratios of the subtracted correlator as well.
The corresponding lines terminate at $n=2$. Again, just like in the case of 
a single resonance (Figs. 2,3,4) the second ratio $r_2$ is affected by the
subtraction. The effect is seen to be smaller at smaller values of the 
fermion mass (higher $\kappa$).
The third ratio $r_3$ is affected very little, while the moment 
ratios for $n>3$ are insensitive to subtractions.

\subsection{Fit of the lower moments}
As explained in the section 1, the correlator of {\em heavy}-quark
currents is special in the way that the lower moments, which originate
from the short-distance
part of that correlator, are determined by perturbative QCD. The lowest 
moments must be saturated by one-loop of free quarks. Therefore we fit those
moments, obtained by Monte-Carlo on the lattice, with one loop of free
Wilson quarks.

In this section we neglect ${\cal O}(\alpha_s)$ terms and obtain an approximate
value of the renormalized heavy-quark mass $\bar{m}_c$ from the
ratio $r_3$:
\begin{equation}
\bar{m}_{c}^{2}\;\;\approx\;\; \frac{1}{4}\,\frac{a_3}{r_{3}^{lat}}   \label{mc_approx}
\end{equation}
We use the coefficient $a_3$ of the continuum theory because the coefficients
$a_n$ of the continuum theory (\ref{rvec_cont}) - (\ref{rsca_cont}) fit
the lattice correlator, composed of Wilson propagators, so well 
(see Fig. 15, 16), provided the mass relation (\ref{m(k)cont}) is used.

Fig. 18 shows the fit of Monte-Carlo data at $n=3$ by the moments of the
pseudoscalar correlator, saturated by one loop of effectively free Wilson
quarks. Those quarks are found to have $\tilde{\kappa}=0.0994$, which for 
free Wilson quarks implies, $a m_{c}^{w} = 1.02$ because of eq.(\ref{m(k)}). 
From the previous section we know that for such a large
mass it is important to correct for the ${\cal O}(a)$ systematic deviation
(\ref{m(k)cont}). One obtains $a \bar{m}_c (a m_{c}^{w} = 1.02) = 0.72$, 
which for $a^{-1} = 1.9\,GeV$ gives:
\begin{equation}
\bar{m}_c\;\;=\;\; 1.37 \,\pm \, 0.01 \; GeV     \label{mc}
\end{equation}
where the error is associated with the contribution of the high-energy part
of the spectrum. 

Other channels, such as vector or scalar, are just as good as the pseudoscalar 
one for the determination of the renormalized quark mass.
The ratio $r_4$ would be also just as good as $r_3$ to determine $\bar{m}_c$, 
since the relation between $r_3$ and $r_4$ is calculable within perturbative QCD.
The difference is $O(\alpha_s)$-effect. We consider it in section (3.4).

Note, that the relation (\ref{m(k)cont}) is derived in the infinite volume. 
Since we 
study rather small lattice $8^3 \ast 16$ we would like to be more accurate with
the finite-volume effects. The moments of the continuum theory (\ref{rpsc_cont})
agree with the lattice moments (see Fig. 15, 16), computed in a rather 
large box $32^4$. In the mean time one can notice in Fig. 14 that the moments
of the small lattices like $8^3 \ast 16$ deviate from the asymptotic 
$L \rightarrow \infty$ behavior even at small $n = \{ 1, 2, 3\}$. In the next 
section we take into account this deviation. 

\subsection{The lattice dispersion relation for the fermionic continuum}
We have seen on Fig. 15, 16 that the moment ratios $r_n$ of the lattice 
correlator in the free-Wilson-quarks approximation follow the corresponding
ratios of the continuum theory with the correct value of the quark mass.
The moments (\ref{rvec_cont}) - (\ref{rsca_cont}) of the continuum theory can be
 calculated with the help of the 
spectral represenation (\ref{mom_disp}) with the spectral density corresponding
to one-loop of free quarks. For the three channels under discussion one can use:
\beqa
\rho^{vec}(s)\;\;&=&\;\;\frac{1}{4\pi^2}
\;\left(s\,+\,2 (\bar{m}_{c})^2\right)\,\sqrt{1\,-\,4 (\bar{m}_{c})^2 / s}
\label{spden_vec}\\
\rho^{psc}(s)\;\;&=&\;\;\frac{3}{8\pi^2}\;s\,\sqrt{1\,-\,4 \bar{m}_{c}/ s}
\label{spden_psc}\\
\rho^{sca}(s)\;\;&=&\;\;\frac{3}{8\pi^2}\;\left(s\,-\,4 (\bar{m}_{c})^2\right)\,
\sqrt{1\,-\,4 (\bar{m}_{c})^2 / s}   \label{spden_sca}
\eeqa
We would like to see how well the lattice variant of the dispersion relation
(\ref{disp_lat}) works for the continuum spectrum of free fermions. 
We can use the spectral densities (\ref{spden_vec}) - (\ref{spden_sca}) in the
eqn. (\ref{disp_lat}) and compare the result with the correlator composed of free
Wilson propagators. We plot the moment ratios obtained in those two ways on
Fig. 17 in the case of small lattice $8^3 \ast 16$. We fit the moments of the
lattice correlator of free Wilson fermions with corresponding moments of the
correlator computed via the dispersion relation (\ref{disp_lat}) with the 
spectral density (\ref{spden_psc}). This is a one-parameter fit. That parameter
is the quark mass of the continuum theory $\bar{m}_{c}$. We find the agreement
illustrated by Fig. 17 remarkable.

Thus, the use of the lattice dispersion relation (\ref{disp_lat}) for the continuum
spectrum of quark-antiquark pair helps to fit the moments on small lattices. It
appeares as an efficient instrument to take into account-finite size effects.
On bigger lattices one can simply use the formula (\ref{m(k)cont}).

\subsection{Fixing the subtraction scheme}
In this section we estimate the value $m_{c}^{\bar{MS}}(m_c)$ of the 
renormalized charmed-quark mass in the
minimal subtraction scheme normalized at the Euclidean point $\mu=m_c$.
In order to fix the subtraction scheme one should be sensitive to the 
$\alpha_s$-correction in the ratios $r_n$. The study of the systematic 
deviations of the coefficient $b_n$ is beyond the framework of the present
paper. However, since the term $b_n \, \alpha_s$ is small in the lower ratios
$r_{2,3,4}$ we will take its values from the continuum theory. This is our
approximation. To be more precise, we write the lower ratios as
\beqa
r_n\;\;&=&\;\;\frac{a_n}{4 m_{c}^{2}}\;\xi_n   \nonumber  \\
\xi \;\,&\equiv&\,\;1\,+\,\alpha_s\,b_n /a_n                   \label{xi} 
\eeqa
We take the value of the strong coupling constant 
$\alpha_s (m_c) \approx 0.3$ 
\footnote{To fix precisely the normalization point of $\alpha_s$ in every 
moment one needs to know the three-loop correction in the ratios (\ref{rqcd}),
which is not known at present}. 
Based on the two-loop calculations of the 
continuum theory \cite{svz1} we obtain for the vector, pseudoscalar and scalar
channels: 
\beqa
\xi_{3}^{psc} = 1.02\;,\;\;\;\;\xi_{3}^{vec} = 0.95\;,\;\;\;\; \xi_{3}^{sca} = 0.97    \nonumber  \\
\xi_{4}^{psc} = 0.96\;,\;\;\;\;\xi_{4}^{vec} = 0.92\;,\;\;\;\;\xi_{4}^{sca} = 0.91    \label{as}
\eeqa
The $\alpha_s$-term is seen to introduce a few-percent correction to the ratios
$r_{2,3,4}$. Therefore, the presently unknown cutoff effects in the coefficient 
$b_n$ have a small effect on the value of the quark mass $\bar{m}_c$. 

We fit the lower moment ratios $r_{3,4}$ of our Monte-Carlo data using the
lattice spectral representation, as explained in the previous section, to
relate more accurately the quark mass of the continuum theory $\bar{m}_{c}$ 
to the corresponding Wilson propagator in the finite-volume lattice theory.

The results for the renormalized charmed-quark mass are shown in the tables.
The tables demonstrate that the dimensionless ratio of the renormalized 
quark mass to the resonance mass $\bar{m}_{c}/m_{res}$ is very stable with 
respect to significant variation of $\kappa$. This implies that the 
renormalized heavy-quark mass is less sensitive to the uncertainty in the 
value of the lattice spacing than one might expect.

The relation $\bar{m}_{c} [\kappa, \beta]$ between the renormalized 
heavy-quark mass and bare parameters of the lattice theory is, in principle,
rather non-trivial. 
On Fig. 18 we plot the moments of one loop of the free Wilson quarks
with the input $\kappa=0.1060$ (dashed line) to show how different they are
from the moments of one loop of free quarks with $\tilde{\kappa}$ 
corresponding to the renormalized mass (bottom line). 
At large values of $\beta$
the relation between $\kappa$ and $\tilde{\kappa}$ is perturbative.
We have generated $20$ gauge configurations for $\beta=20$ with the same
Monte-Carlo parameters as for $\beta=6$. The corresponding moments of the
correlator with the clover-and-tadpole-improved propagators for the same
input $\kappa=0.1060$ are shown on Fig. 18 by squares. The Monte-Carlo data
for $\beta=20$ is fitted perfectly at all $n$ by one loop of free-quarks 
(upper solid line) with $\kappa ' = 0.1034$. We find numerically the 
relation :
\bb
\kappa ' = \kappa <U_{\Box}(\beta=20)>^{1/4}     \label{beta20}
\ee
Obviously the data at $\beta=20$ shows the correlator at distances so small 
that the effects due to interaction are not seen. The contribution of the 
clover term is strongly suppressed. The relation (\ref{beta20}) is due to 
the fact that we used the mean-field-tadpole-improved action \cite{lm}. 
Generating the correlator by Monte-Carlo at various high values of $\beta$ 
one can study the correlator in the weak-coupling expansion \cite{bdf}.

\section{Conclusions}
We have formulated in details the way to calculate the renormalized quark 
mass using Monte-Carlo data on the correlator of heavy-quark currents.
We separate the long-distance and the short-distance parts of the
correlator by studying the higher and the lower moments
of the correlator respectively. The long-distance
part is used to fix the scale of the lattice theory, as it is saturated
by the lowlying resonance. The short-distance part is fitted by the
loop expansion of perturbative QCD. The coefficients of that expansion
are contaminated by the cutoff effects and the finite-volume effects.
Those systematic deviations can be studied order by order in the
loop-expansion. We have worked out systematic deviations of the one loop
of Wilson quarks, which saturates the very short-distance part of the
correlator (the lowest moments). Then fit to Monte-Carlo data
produces the value of the renormalized heavy-quark mass. We find for
the charmed quark mass: $m_{c}^{\bar{MS}}(m_c)\,=\,1.22(2) GeV$ - from
the vector channel and $m_{c}^{\bar{MS}}(m_c)\,=\,1.18(2) GeV$- from
the pseudoscalar channel. There are two sources of uncertainties in these
values. The first is the ansence of scaling on the present lattice. The data
for higher $\beta > 6$ as well as improved actions should helpful in this 
respect. The second uncertainty is in fixing the scale. One needs bigger 
volume to determine the resonance mass with better accuracy. 
The dimensionless ratio of the renormalized heavy-quark mass to the resonance 
mass is stable with respect to variation of $\kappa$ provided the value
of $\kappa$ corresponds to heavy quarks.

Our central value for the charmed quark mass in the vector channel can be
compared to the corresponding estimate of the continuum theory : 
$m_{c}^{\bar{MS}}(m_c)\,=\,1.23 GeV$ \cite{nar}. There are no estimates 
of the charmed quark mass from the pseudoscalar or scalar correlators in the
continuum theory, since the corresponding spectral densities are not 
observable experimentally.

A comprehensive study of the correlator of heavy-quark currents at both
large and small distances is a challenging problem.
One principal question we tried to clarify here is whether one can observe
simultaneously, on
the lattice sizes presently available to Monte-Carlo simulations, both
short-distance phenomena, described by asymptotic freedom, and
long-distance tails of hadronic correlators, saturated by low-lying
resonances. In order to see short-distance phenomena one has to keep the
lattice spacing small. Then one needs many lattice sites in order to reach
large physical scales, where the low-energy part of the hadronic spectrum
dominates. To solve this problem it is crucial to combine the extension
of dispersion relations to the lattice with the improved actions.
The former helps to fit correlators in smaller volumes, while the latter
helps to see QCD perturbation theory on coarse lattices.
Further study of the $\alpha_s$-corrections \cite{bdf} to the correlator of the
heavy-quark currents on the lattice is necessary to specify the renormalization 
scheme of the extracted quark mass more accurately and measure the renormalized 
strong coupling constant.

\section*{Acknowledgments}
One of the authors (A. B.) would like to thank P. van Baal, P. Lepage,
L. McLerran, J. Kapusta, A. Kovner, M. Shifman, J. Smit, A. Vainshtein,
M. Voloshin, R. Willey for useful discussions and DOE for support under
the grant DE-FG02-87ER40382. 
We thank the QCD-TARO collaboration for generating the blocked $SU(3)$
configurations we used.
Computer time for this project was provided by the Minnesota
Supercomputer Institute and by the Pittsburgh Supercomputer Center.

\newpage
\begin{center}
\begin{tabular}{|c|c|c|c|c|c|}
\hline 
fit to $r_3$&$ m_{res} $&$ \bar{m}_{c} $&$ \bar{m}_{c}/m_{res}
$&$  m_{c}^{\bar{MS}}[a_{J/\psi}]$
&$ m_{c}^{\bar{MS}}[a_{\eta_c}]$\\ \hline
$\eta_c$ & 2.024(9) & .779(2) & .385 &  1.18 & 1.16\\
$J/\psi$ & 2.06(1)  & .835(2) & .406 &  1.23 & 1.20\\
$\chi_o$ & 2.29(12) & .778(9) & .340 &  1.15 & 1.13\\ \hline
fit to $r_4$&$ m_{res} $&$ \bar{m}_{c} $&$ \bar{m}_{c}/m_{res}
$&$ m_{c}^{\bar{MS}}[a_{J/\psi}]$
&$ m_{c}^{\bar{MS}}[a_{\eta_c}]$\\ \hline
$\eta_c$ & 2.024(9) & .815(2)  & .403 &  1.20 & 1.18\\
$J/\psi$ & 2.06(1)  & .861(3)  & .418 &  1.24 & 1.22\\
$\chi_o$ & 2.29(12) & .827(10) & .361 &  1.19 & 1.16\\ \hline 
\end{tabular}
\end{center}
\vspace*{3cm}
Table 1. Charm-quark mass as a result of the fit of the ratios $r_3$, 
$r_4$ of the subtracted correlator for $\kappa = 0.1000$. The masses $m_{res}$
and $\bar{m}_{c}$ are in units of the lattice spacing. The masses 
$m_{c}^{\bar{MS}}[a_{J/\psi}]$ and $m_{c}^{\bar{MS}}[a_{\eta_c}]$ are in $GeV$. 
They are obtained assuming that the lattice spacing is fixed 
from the vector channel and from the pseudoscalar channel respectively. 
The indicated errors are statistical.
\newpage
\begin{center}
\begin{tabular}{|c|c|c|c|c|c|}
\hline 
fit to $r_3$&$ m_{res} $&$ \bar{m}_{c} $&$ \bar{m}_{c}/m_{res}
$&$  m_{c}^{\bar{MS}}[a_{J/\psi}]$
&$ m_{c}^{\bar{MS}}[a_{\eta_c}]$\\ \hline
$\eta_c$ & 1.565(1) & .596(2) & .381 &  1.16 & 1.15 \\
$J/\psi$ & 1.61(2)  & .651(4) & .404 &  1.22 & 1.21 \\
$\chi_o$ & 1.87(10) & .609(9) & .326 &  1.15 & 1.14 \\ \hline
fit to $r_4$&$ m_{res} $&$ \bar{m}_{c}/m_{res}
$&$ \bar{m}_{c}$&$ m_{c}^{\bar{MS}}[a_{J/\psi}]$
&$ m_{c}^{\bar{MS}}[a_{\eta_c}]$\\ \hline
$\eta_c$ & 1.57(1)  & .623(3) & .398 &  1.17 & 1.16 \\
$J/\psi$ & 1.61(2)  & .666(4) & .414 &  1.23 & 1.22 \\
$\chi_o$ & 1.87(10) & .649(10)& .347 &  1.19 & 1.18 \\ \hline 
\end{tabular}
\end{center}
\vspace*{3cm}
\begin{center}
{Table 2. The same as in Table 1 for $\kappa = 0.1060$}.
\end{center}

\end{document}